\documentclass[aps,twocolumn,prc,floatfix,showpacs,preprintnumbers,amsmath,amssymb,nofootinbib,groupedaddress]{revtex4-1}
\usepackage[normalem]{ulem}
\usepackage{wasysym}
\usepackage{color}
\usepackage{graphicx}
\usepackage{dcolumn}    
\usepackage{bigstrut}
\usepackage{amsmath,amsfonts,amsthm,bm}
\usepackage{bm}
\usepackage{CJK}

\def\be{\begin{equation}}
\def\ee{\end{equation}}
\def\bea{\begin{eqnarray}}
\def\eea{\end{eqnarray}}
\def\bse{\begin{subequations}}
\def\ese{\end{subequations}}
\def\br{{\bf r}}

\begin{document}
%
\title{Comment on ``Breakdown of the tensor component in the
Skyrme energy density functional''}


\author{H. Sagawa}
\email{sagawa@ribf.riken.jp}
\affiliation{RIKEN Nishina Center, Wako 351-0198, Japan}
\affiliation{Center for Mathematics and Physics, University of Aizu, Aizu-Wakamatsu, Fukushima 965-8560, Japan}

\author{G. Col\`o}
\email{gianluca.colo@mi.infn.it}
\affiliation{Dipartimento di Fisica ``Aldo Pontremoli'', Universit\`a degli Studi di Milano, 20133 Milano, Italy}
\affiliation{INFN, Sezione di Milano, 20133 Milano, Italy}

\author{Ligang Cao}
\email{caolg@bnu.edu.cn}
\affiliation{College of Nuclear Science and Technology, Beijing Normal University, Beijing 100875, China}

\date{\today}

\begin{abstract}
In a recent paper [Phys. Rev. C 101, 014305 (2020)], Dong and Shang 
claim that the Skyrme original tensor interaction is
invalid. Their conclusion is based on the misconception that the Fourier transform of 
tensor interaction is difficult or even impossible, so that the Skrme-type tensor 
interaction was introduced in an unreasonable way. We disagree on their claim. In this note, we show that one can easily get the Skyrme force in momentum space by Fourier transformation if one starts from a general central, spin-orbit or tensor interaction with
a radial dependence.
\end{abstract}


\maketitle

\section{Introduction}

In a recent paper published in Phys. Rev. C (Ref. \cite{Dong2020}, hereafter referred to as
Dong and Shang), the authors claim in the abstract that ``the Skyrme original tensor interaction [...] is
invalid.'' One has to remind that, although the original idea
by Skyrme dates back to long time ago and the Skyrme tensor force has been written out in
Refs. \cite{Rehovoth,Skyrme1}, there have been many studies on how to implement it and fit
its parameters since then, even in recent years.
A first series of papers appeared in the 1970s, when the first
Skyrme forces had been proposed as a practical tool for Hartree-Fock (HF) and Random Phase
Approximation (RPA). A second series, much more numerous, has appeared in the years since 2000.
In 2014, some of us published a review paper on the tensor force within mean-field and
Density Functional Theory (DFT) approaches to nuclear structure \cite{review}. In that review,
we quoted $\approx$30-40 papers where the Skyrme tensor is implemented and studied, authored by
different colleagues. The conclusion of that review paper is that evidences for a strong
neutron-proton tensor force exist, even in mean-field and DFT studies, while the role of
the tensor force between equal particles (neutron-neutron or proton-proton) is less well
established albeit not completely ruled out. Therefore, if the tensor force proposed by
Skyrme were ``invalid'', this would impact on a lot of published works and conclusions drawn so far.
Accordingly, it would be appropriate to have a strong argument regarding this ``invalidity''.

The argument against the Skyrme tensor force should be found
in Sec. II of the paper. In this Section, one can read: ``The Skyrme original tensor force was
introduced in an unreasonable way, because the tensor-force operator $S_{12}$ in momentum space
but with an $r$-dependent strength, i.e., $f_T(r)S_{12}(\mathbf k)$, is applied as a starting point.''
We remind, for the reader's convenience, that
\be\label{eq:tensor}
S_{12}(\mathbf k)=(\sigma_1\cdot\mathbf k)(\sigma_2\cdot \mathbf k)-{k^2}\frac{\sigma_1\cdot\sigma_2}{3}.
\ee
We cannot see any logic behind the latter sentence in the work by Dong and Shang. In fact, it does not
have any formal ground. In physics, nothing forbids an interaction to be at the same time
position-dependent and momentum-dependent. 
Bethe \cite{Bethe} was one of the first to advocate that this ought to
be the case if one wishes to introduce an effective potential for finite nuclei. Skyrme echoed this
at the beginning of his first paper on this topic, and it is useful at this stage to quote literally the
sentence from Ref. \cite{Skyrme}: ``in the case of a finite system the effective potential must depend
upon both momenta and coordinate.''

In fact, this latter sentence does not simply lie as a ground for the Skyrme tensor force, but rather
it is at the basis of the whole philosophy behind Skyrme forces and Skyrme functionals \cite{Skyrme,Bell}.
Several authors have proposed Skyrme-type forces with terms that have both momentum- and density-dependence.
If the Skyrme tensor force is unreasonable because of the reason advocated by Dong and Shang,
the whole Skyrme force will be unreasonable. If this were the case, this would disgrace not 
only the results
of a few hundred papers in which Skyrme tensor terms are introduced, but also some more
$\approx 10^3$ papers in which Skyrme forces are used, including the paper by Dong and Shang themselves.
As we said, we see no reason to rule out a force because it looks like a momentum-dependent operator
times a function of the relative coordinate $r$.

It should also be noticed that any comparison with a realistic potential used in Br\"uckner-HF
calculations, or similar cases, is immaterial in this context. We are not discussing any
{\em realistic} potential but only the case of an {\em effective} potential to be used in
HF or DFT. Skyrme himself, in Ref. \cite{Rehovoth}, employed the word ``pseudopotential''
to distinguish clearly his approach from any one based on a fundamental interaction.
In modern language, we would say that we deal with an effective feld theory in which
the coupling terms depend on $\delta(\bf r_1-\bf r_2)$ times derivatives. An analogous
case is that of the pionless EFT.

In the paper by Dong and Shang, it is also claimed that the Skrme-type tensor interaction is introduced
in an unreasonable way since the Fourier transform of such tensor interaction is difficult or even impossible.
We disagree on this claim based on the procedure
that we discuss explicitly in the next Section.
In the next Section we will show that if we start from a general central, spin-orbit or tensor interaction, with
a radial dependence such that the range is very short, the Fourier transform produces the Skyrme force in momentum space.
This is a further, more detailed and mathematically rigorous, way to show that the arguments in the paper
by Dong and Shang are invalid. 

\section{Fourier transform of the terms of the Skyrme force}

Let us write the Fourier transform of an interaction $V(r)$ as
\be \label{eq1}
V({\bf q})=\int e^{i{\bf q}\cdot \br}V(r) d\br,
\ee
where ${\bf q} ={\bf k}-{\bf k}'$
is the momentum transfer (${\bf k}$ and ${\bf k}'$ are the initial and final relative momenta).
For a central interaction $V_C(r)$, the integral in \eqref{eq1} can be performed
by using the multipole expansion of a plane wave
\be\label{eq2}
e^{i{\bf q}\cdot \br}=4\pi\sum_{\lambda\mu}i^{\lambda}j_\lambda(qr)Y_{\lambda\mu}(\hat{q})Y^*_{\lambda\mu}(\hat{r}),
\ee
where $j_\lambda(qr)$ is the spherical Bessel function of rank $\lambda$, and $Y_{\lambda\mu}(q)$ is a spherical harmonic.
Thus, the integral for a central interaction becomes 
\be\label{eq3}
V({\bf q})\propto \int j_0(qr)V_C(r)r^2dr.
\ee
The constant and the lowest-order momentum-dependent terms of the Skyrme interaction are obtained by means of the
Taylor expansion $j_0(qr)=1-(qr)^2/2+\cdots$.
The constant term of this expansion provides the $t_0$ term in the Skyrme force.
The second term is written as
\be \label{eq3a}
V({\bf q})\propto {\bf q}^2\int V_C(r)r^4dr,
\ee
with
\be\label{eq4}
{\bf q}^2={\bf k}^2+{\bf k'}^2-2{\bf k}\cdot{\bf k'}.
\ee
The first two terms in Eq. \eqref{eq4} provide the $t_1$ term of the Skyrme force, while the third term gives the $t_2$ term. These
two terms
mimic the finite range in the effective two-body interaction.
The radial integrals in Eqs. \eqref{eq3} and \eqref{eq3a} can be easily performed for any commonly adopted Yukawa-type or
Gaussian-type finite-range interaction.

The spin-orbit interaction $V_{LS}$,
\be\label{eq5}
V_{LS}=f_{LS}(r)(\sigma_1+\sigma_2)\cdot (\br_1-\br_2)\times ({\bf p}_1-{\bf p}_2),
\ee
can be also expanded, after being Fourier-transformed through Eq. \eqref{eq1}, in the momentum operator $\bf q$.
The radial dependence of $V_{LS}(r)$ is expressed by the spherical harmonics $Y_{1\mu}$,
so that the expansion of Eq. \eqref{eq2} produces the spherical Bessel function $j_1(qr)$. We do not repeat these steps here
as they can be found in Ref. \cite{Bell}.

The tensor interaction in the coordinate space is expressed as
\be\label{eq6}
V_T(r)=f_T(r)S_{12}(\br),
\ee
where
\be\label{eq7}
S_{12}(\br)=(\sigma_1\cdot\br)(\sigma_2\cdot \br)-{r^2}\frac{\sigma_1\cdot\sigma_2}{3}.
\ee
The tensor operator of \eqref{eq7} is rewritten using the spherical harmonic $Y_{2\mu}$ and becomes
\be
S_{12}(\br)=\sqrt{\frac{8\pi}{3}}[(\sigma_1\cdot\sigma_2)^{(2)}\times r^2Y_2(\hat r)]^{(0)}.
\ee
The Fourier transform can be evaluated as
\be
V_T({\bf q})=\int  e^{i{\bf q}\cdot \br}V_T(r) d\br \propto \int r^4dr j_2(qr)[(\sigma_1\cdot\sigma_2)^{(2)}\times Y_2(\hat q)]^{(0)}.
\ee
The spherical Bessel function $j_2(qr)$ is proportional to $q^2$, as $ j_2(qr)\sim (qr)^2/5!!$ in the lowest order of
expansion, and $V_T(q)$ becomes
\be
 V_T({\bf q})\simeq -\frac{4\pi}{15}\sqrt{\frac{8\pi}{3}}[(\sigma_1\cdot\sigma_2)^{(2)}\times q^2Y_2(\hat q)]^{(0)}\int r^6f_T(r) dr.
\ee
In this way, we obtain the tensor interactions in momentum space in the form
\bea
S_{12}({\bf q}) &=& \sqrt{\frac{8\pi}{3}}[(\sigma_1\cdot\sigma_2)^{(2)}\times q^2Y_2(\hat q)]^{(0)} \\
&=&(\sigma_1\cdot{\bf q})(\sigma_2\cdot {\bf q})-{\bf q}^2\frac{\sigma_1\cdot\sigma_2}{3} \\
&=&\{[(\mathbf{\sigma _{1}\cdot {k}^{\prime})(\sigma _{2}\cdot {k}^{\prime })-\frac{1}{3}\left( \sigma_{1}\cdot \sigma _{2}\right) {k}^{\prime 2}] }\nonumber\\
&+&[ (\bf{\sigma _{1}\cdot {k})(\sigma _{2}\cdot
{k})-\frac{1}{3}\left(
\sigma _{1}\cdot \sigma _{2}\right) {k}^{2}}] \}\nonumber\\
&-&\{\left( \sigma _{1}\cdot \bf{k}^{\prime }\right)
(\sigma _{2}\cdot \bf{k}) +\left( \sigma _{2}\cdot
\bf{k}^{\prime }\right) (\sigma _{1}\cdot
{k})\nonumber\\
&-&\frac{2}{3}\left[ (\bf{\sigma }_{1}\cdot \bf{\sigma }_{2})
\bf{k}^{\prime }\cdot \bf{k} \right] \}.\label{tensor}
\eea
In the above expression, the operator ${\bf k}=\left(\bf
\nabla_1-\bf \nabla_2\right)/2i$ acts on the right and ${\bf
k}^\prime=-\left(\bf \nabla_1^{\prime}-\bf
\nabla_2^{\prime}\right)/2i$ acts on the left. The first two terms in Eq. \eqref{tensor} are the so-called triplet-even tensor term
(T-term in Skyrme), whereas the third and fourth terms correspond to a triplet-odd term (U-term).
In this way we can validate the Skyrme type tensor interaction as the lowest-order expansion of a finite-range
tensor force.

\end{document}